\documentclass[prl,aps,twocolumn,superscriptaddress, showpacs]{revtex4}
 \pdfoutput=1
\usepackage{graphicx}
\usepackage{epsfig}
\usepackage{hyperref}
\usepackage{dcolumn}
\usepackage{amssymb}
\usepackage{amsmath}
\usepackage{natbib}

\newcommand{\Real}[1]{\textsf{Re}(#1)}

\usepackage{amsmath}
\usepackage{bbm,amsfonts}

\def\E{{\mathcal E}}

\usepackage{amsthm}
\theoremstyle{definition}

\input{Qcircuit.tex}
\newcommand{\beqa}{\begin{eqnarray}}
\newcommand{\eeqa}{\end{eqnarray}}
\newcommand{\beq}{\begin{equation}}
\newcommand{\eeq}{\end{equation}}

\begin{document}
\title{Selective and efficient quantum process tomography with single photons}

\author{Christian Tom\'as~Schmiegelow}
\affiliation{Departamento de F\'{\i}sica \& IFIBA, 
FCEyN, UBA, Pabell\'on 1, Ciudad Universitaria, 1428 Buenos Aires,
 Argentina}

\author{Miguel Antonio~Larotonda}
\affiliation{CEILAP, CITEDEF, J.B. de La Salle 4397, 1603 Villa Martelli,
Buenos Aires, Argentina}

\author{Juan Pablo~Paz}
\affiliation{Departamento de F\'{\i}sica \& IFIBA, 
FCEyN, UBA, Pabell\'on 1, Ciudad Universitaria, 1428 Buenos Aires,
 Argentina}


\begin{abstract}
We present the results of the first photonic implementation of a new method for quantum process tomography. The method (originally presented by A. Bendersky et al, Phys. Rev. Lett {\bf 100}, 190403 (2008)) enables the estimation of any element of the chi-matrix that characterizes a quantum process using resources that scale polynomially with the number of qubits. It is based on the idea of mapping the estimation of any chi-matrix element onto the average fidelity of a quantum channel and estimating the latter by sampling randomly over a special set of states called a 2-design. With a heralded single photon source we fully implement such algorithm and perform process tomography on a number of channels affecting the polarization qubit. The method is compared with other existing ones and its advantages are discussed. 
\end{abstract}


\maketitle

The complete characterization of a linear quantum process mapping initial states into final states ($\E(\rho_{in})=\rho_{out}$) is a very hard task. In fact this well known fact can be understood as follows: A general completely positive (CP) map can be fully characterized by the so-called $\chi$--matrix\cite{NC}. Choosing the basis $E_a$ consisting of $D^2$ operators, the $\chi$-matrix representation for the map $\E$ is such that $\E(\rho) = \sum_{ab} \chi_{ab}E_a \rho E_b^\dagger$ ($D=2^n$ for a system of $n$ qubits). The matrix $\chi_{ab}$ must be positive and hermitian for the channel to be completely positive (CP) and hermitian itself. Moreover, for the map to be trace preserving its $\chi$--matrix should be such that the condition $\sum_{ab} \chi_{ab} E_a^\dagger E_b = I$ is satisfied. The matrix $\chi_{ab}$ is defined by $D^4-D^2$ real parameters, a number that scales exponentially with the number of qubits of the system. This implies that full quantum process tomography (QPT) is unavoidably hard.  Moreover, until recently, known methods to estimate matrix elements $\chi_{ab}$ were also inefficient since they require resources (measured in terms of the number of repetitions of each experiment, on the number of operations required to perform them, etc) scaling exponentially with the number of qubits \cite{NC,mohseni20067direct,d2001quantum}. Recently we introduced an efficient strategy enabling the estimation of any $\chi_{ab}$ element investing for such purpose resources scaling polynomially with the number of qubits of the system (and on the accuracy required for the estimation) \cite{bendersky20089selective}. In this paper we will present the first successful  experimental implementation of such strategy using single photons and linear optical elements. The experiment is not only an illustration of the use of a general scheme but also makes clear the advantages of the new tomographic method over its predecessors. 

It is convenient first to present the main ingredients of the only available method to perform selective and efficient quantum process tomography (SEQPT). For this purpose, it is best to first notice a general property of quantum channels. Assuming only that the operator base $E_a$ is orthonormal in the Hilbert Schmidt inner product, it can be shown that any element $\chi_{ab}$ is related to the average fidelity of a modified channel $\E_{ab}$. Such channel depends on the original $\E$ and on the operators $E_a$ and $E_b$. It acts on any state $\rho$ as $\E_{ab}(\rho)=\E(E_a \rho E_b)$. Thus, we can show that  \cite{bendersky20089selective}
\beq
\chi_{ab}= {1\over D} \left(
(D+1)F(\E_{ab})-\delta_{ab}\right).
\eeq 
Here, the average fidelity $F(\E_{ab})$ is computed integrating over all pure states $|\phi\rangle$ using the (unitarily invariant) Haar measure, i.e. 
\beq
F(\E_{ab})=\int d|\phi\rangle 
\langle\phi| \E_{ab}(|\phi\rangle\langle\phi|) |\phi\rangle. 
\eeq 
The method of SEQPT is based on the use of the above identity and on the fact that the integral over the entire Hilbert space can be exactly computed using finite resources. It has been shown that the average fidelity can be exactly evaluated by computing the average over a finite set of states that form a so-called $2$--design \cite{delsarte1977spherical}. In fact, if the set $S=\{|\psi_j\rangle, j=1,...,K\}$ is a $2$--design, we have
\beq
F(\E_{ab}) = \frac{1}{K}\sum_j \langle\psi_j| \E_{ab}(|\psi_j\rangle\langle\psi_j|) |\psi_j\rangle.
\eeq 
Although the exact computation of $\chi_{ab}$ using a 2--design would involve performing a sum over an exponentially large set, the estimation of such coefficients with finite ($D$--independent) precision is possible by randomly sampling over states  $|\psi_j\rangle$. As described in \cite{bendersky20089selective} and discussed with more detail below, the precision in the estimation scales better than $1/\sqrt{M}$ with the number of repetitions $M$ of the experiment.

There are several known examples of $2$--designs such as the one formed by the $D(D+1)$ states  belonging to $D+1$ mutually unbiased bases\cite{klappenecker2005mutually} (two bases are mutually unbiased iff the absolute value of the overlap between states in such bases is $1/\sqrt{D}$). The equation (3) suggests an immediate way to characterize the quantum channel: For every coefficient $\chi_{ab}$ one needs to compute the survival probability of states $|\psi_j\rangle$ when evolved over the channel $\E_{ab}$ and then average over the $2$--design. For the strategy to be viable one needs not only to make sure that the channel can be implemented efficiently but also that the sum over the elements of the $2$--design can be estimated with polynomial resources. The fact that these two tasks can be performed is the core of SEQPT. 
In particular, the diagonal elements $\chi_{aa}$ are average fidelities of the CP-map $\E_{aa}$, obtained by composing the original map $\E$ with the unitary operation $E_a$: $\E_{aa}(\rho)=\E(E_a\rho E_a)$.  Evaluating off diagonal elements $\chi_{ab}$ requires a different strategy that at first sight seems to be rather different than the one used for diagonal coefficients. However, the main point in this strategy, again, is to realize that any off diagonal coefficient $\chi_{ab}$ is related through (1) with the average fidelity of the map $\E_{ab}$. This map is not CP but can be obtained as the difference between two CP maps. For this reason, as shown in \cite{bendersky20089selective}, the off diagonal coefficients can be obtained as the mean value of an ancillary qubit conditioned to the survival of the state $\ket{\psi_j}$ and averaged over the 2--design. The complete algorithm is described in detail in Figure 1. In fact, real and imaginary parts of $\chi_{ab}$ are obtained by conditionally measuring the mean values of $\sigma_x$ and $\sigma_y$ of an ancillary qubit that interacts with the system with controlled--$E_{a,b}$ operations. 

\begin{figure}[ht]
\begin{equation*}
\Qcircuit @C=.9em @R=.4em {
     \ket{0}_{\text{Ancilla}}&& & & \gate{H} &
         \ctrl{1} & \ctrlo{1} &
        \qw & \meter & \sigma_x
     \\
     \ket{\psi_j}_{\text{Main}}&& & & {/}\qw &   
         \gate{E_a^\dagger} & \gate{E_{b}^\dagger} & \gate{\mbox{\huge $\E$}} &
          \meter & ~\Pi_{\psi_j}
\gategroup{1}{5}{2}{8}{.7em}{--}
}
\end{equation*}
\caption{The quantum algorithm for measuring $\Real{\chi_{ab}}$ for a given channel $\E$. The method requires an extra ancillary (clean) qubit. The imaginary part of $\chi_{ab}$ is estimated in the same way by measuring the polarization of the ancilla along the $y$ axis.}
\label{circ:offdiag} 
\end{figure}
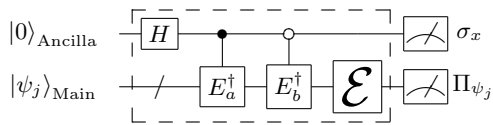

Although SEQPT is the only available scheme to efficiently estimate any coefficient of the chi-matrix it has never been implemented in practice. Other methods have been demostrated in different experimental setups (see, for example, \cite{emerson2007symmetrized,liu2008direct,monz2009realization, altepeter2003ancilla,la_ceci, limblad}). We present here the results of the first implementation of SEQPT in a photonic quantum information processor. For this we use a heralded single photon source encoding two qubits in a single photon. We do this by using both the polarization degree of freedom and the momentum (path) degree of freedom of a heralded photon generated by parametric-down-conversion  \cite{englert2001universal}. The method enables us to characterize unknown quantum channels affecting the polarization qubit using the path qubit as an ancilla.

\begin{figure}[t]
\begin{center}
    \epsfig{file=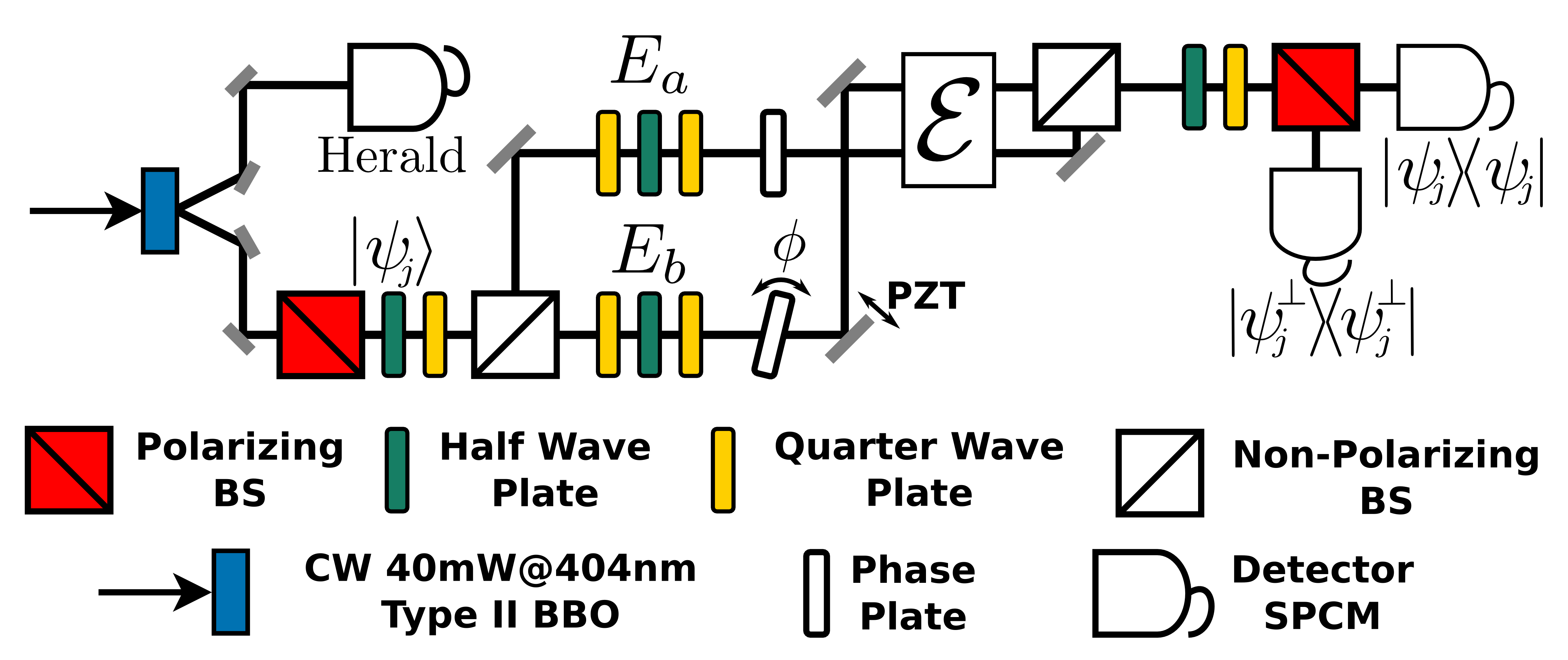, angle=0, width=0.47\textwidth}
\end{center}
\vspace{-15pt} 
\caption{A continuous laser diode at 405nm and 40mW generates frequency degenerate twin photons on BBO crystal, cut for Type-II parametric-downconversion. One photon is used as a herald while on the other a polarization and path qubit are encoded.  We first prepare a state from the 2--design on the polarization qubit with a polarizer, half and quarter wave-plates, then we perform a Hadamard gate on the path with a non-polarizing 50:50 beam-splitter, next we perform the controlled operations $E_a$ and $E_b$ with a quarter-half-quarter wave plate sequence on each arm, both paths are then sent through the unknown channel $\mathcal{E}$ affecting the polarization qubit equally on both paths. Finally measurement of $\sigma_x$ or $\sigma_y$ in the path qubit is done by interfering both paths at a second non-polarizing beam-splitter with a phase $\phi$ previously set by tilting a glass; projection on the prepared polarized state is done by a quarter and half wave-plate followed by a polarizing beam-splitter and detection is done by fiber coupled avalanche photodiodes. The interferometer has an average visibility of $92\%$. Not shown in the figure is an intense beam counter-propagating in the interferometer which is monitored to actively control the position of one of its mirrors with a piezoelectric disc attaining a stability of over $\lambda/30$ over all measurements.}
\label{fig:scheme}
\end{figure}

The full implementation is schematically shown in Figure \ref{fig:scheme}. For the sake of clarity we divide the description of the method in four steps: i) generation of initial states of the $2$--design $|\psi_j\rangle$,  ii) application of controlled operations between ancillary qubit and target qubit , iii) the evolution through the unknown process and  iv) measurement in the $\sigma_x$ or $\sigma_y$ base for the ancilla (path) qubit conditioned on the survival of the $|\psi_j\rangle$ (polarization) states.  With these steps we implement the algorithm of Figure \ref{circ:offdiag} is with the photonic setup described in Figure \ref{fig:scheme}.
Stage (i) is rather simple. For the case of a single qubit it turns out that the $2$--design is formed by the $D(D+1)=6$ eigenstates of the three Pauli operators (which define three mutually unbiased bases). For the polarization of photons these are: two states with vertical-horizontal polarization, two with diagonal (45-135 degrees) polarization and two with circular (right-left) polarization. The generation of such states is done with the standard polarizer, half and quarter wave plate configuration. Stage (ii) of the algorithm is conceptually simple: the photon is split by the first non-polarizing beam-splitter, which acts as the Hadamard gate in the ancillary qubit, then any controlled operation on the polarization qubit can done by rotating independently the polarization in each path using a quarter-half-quarter wave plate configuration\cite{englert2001universal}.  In order to apply the unknown process (iii) both paths are then made parallel and sent though a zone where the unknown channel $\E$ is performed on the polarization qubit. Finally (iv) to measure the expectation values of $\sigma_x$ and $\sigma_y$ conditioned to the survival of $|\psi_j\rangle$ we make both paths interfere with a relative phase $\phi$ at a second non-polarizing beam-splitter by tilting a glass in one of the paths and project into the polarization states with the inverse quarter, half wave plate and polarizer configuration as in (i). With this scheme measure the quantities $p_{ab}(\sigma_{x(y)},\pm;\Pi_{\psi_j})$, which are the probabilities of finding the ancilla in the $\pm$ state of the $x(y)$ base conditioned on the survival of the input states $\ket{\psi_j}$ for each $\E_{ab}$. With these probabilities we obtain all the necessary data to determine the matrix element $\chi_{ab}$.  A note must be made about diagonal elements: as mentioned above, in this case the scheme simplifies significantly because there is no need for an ancilla qubit so one then can look at only one arm of the interferometer by blocking the other one.

\begin{figure}[h]
\begin{center}
    \epsfig{file=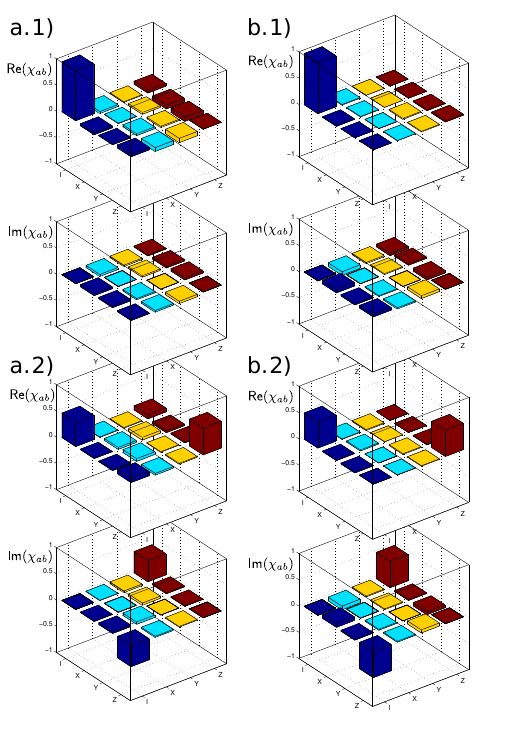, angle=0, width=0.4\textwidth}
\end{center}
\vspace{-15pt} 
\caption{Experimental results for SEQPT (a.1) and (a.2) and Standard QPT (b.1) and (b.2). First two rows display measured real (top) and imaginary (bottom) parts of the matrix $\chi_{ab}$ of the quantum process corresponding to free propagation (identity channel). Similarly the last two rows display correspond to real (top) and imaginary (bottom) parts
of the matrix $\chi_{ab}$ measured for a process corresponding to a QWP at $0^\circ$.}
\label{fig:results}
\end{figure}

The elements of $\chi_{ab}$ were detected for two different noisy processes affecting the polarization degree of freedom. The results are shown in Figure \ref{fig:results}. We must stress that although we measured all the matrix elements characterizing both channels our method determines any $\chi_{ab}$ independently and efficiently (i.e., as opposed to previous methods \cite{NC} it is not necessary to fully characterize the channel in order to determine a single $\chi_{ab}$ coefficient). The measured processes correspond to the identity channel (i.e., free propagation through air) and to a quarter wave plate at $0^\circ$. Those processes were also fully characterized by means of the standard method of QPT (as explained in chapter 10 of \cite{NC}). The results obtained by both methods are compared in Figure \ref{fig:results} and turn out to be in very good agreement. As a figure of merit to compare both schemes we numerically calculate fidelity between the channels determined by each method obtaining  $F=95,1\%\pm1.5\%$ for the identity and $F=96,3\%\pm1.6\%$ for the half wave-plate at $0^\circ$ \footnote{The fidelity between channels is commonly defined as $F[\mathcal{E}_1, \mathcal{E}_2]=\int f(\mathcal{E}_1(\rho),\mathcal{E}_2(\rho)) d \ket{\psi}$ where $\rho=\ket{\psi}\!\bra{\psi}$ and $f[\rho_1,\rho_2]=\left( Tr (\sqrt{\! \sqrt{\rho_1}\rho_2 \sqrt{\rho_1}  }) \right)^2$, see \cite{altepeter2003ancilla}}. The maximum fidelity attainable is mainly limited by interferometer visibility, which results in a not perfectly clean ancilla. 

SEQPT is a method that is suited to perform partial process tomography selecting the relevant parameters one wants to estimate and investing polynomial resources for such estimation. It is interesting to compare the resources required to implement this new tomographic method and previously existing ones. To determine any single matrix element $\chi_{ab}$ using SEQPT, we measured $D(D+1)=6$ survival probabilities. On the contrary, obtaining a single matrix element $\chi_{ab}$ using the standard QPT \cite{NC} requires estimating $D^2\!\times\!D^2=16$ transition probabilities. This implies that even at the level of a single qubit the SEQPT is more "efficient". This comparison might seem somewhat unfair since after such number of experiment the standard method provides all the information required to estimate the full $\chi_{ab}$ matrix and not just  a single element of such matrix.  However, when working with larger systems, it is clear that one would never estimate the exact matrix elements $\chi_{ab}$ by performing the average over the entire $2$--design. In fact, this would always be exponentially hard (as the $2$--design is a set containing an exponentially large number of elements). Instead, the main virtue of SEQPT is that it enables us to estimate any element with fixed accuracy by performing a number of measurements that only depends on such accuracy and is independent on the number of qubits. As shown in \cite{bendersky20089selective} the error in the estimation of the average in equation (3)  after $M$ experiments scales as $1/\sqrt{M}$. However, it is interesting to get a tighter bound for such error. In fact, if the average fidelity is computed by sampling over states $|\psi_j\rangle$ which are never prepared twice, then one can show that the error in the average turns out to scale as $\Delta\chi\propto\sqrt{\frac{1}{M}(1-\frac{M-1}{K-1})}$ where $K=D(D+1)$ is the number of elements in the $2$--design. In fact, this equation tells us that the error vanishes as the number of experiments $M=K$. 

\begin{figure}[t]
\begin{center}
    \epsfig{file=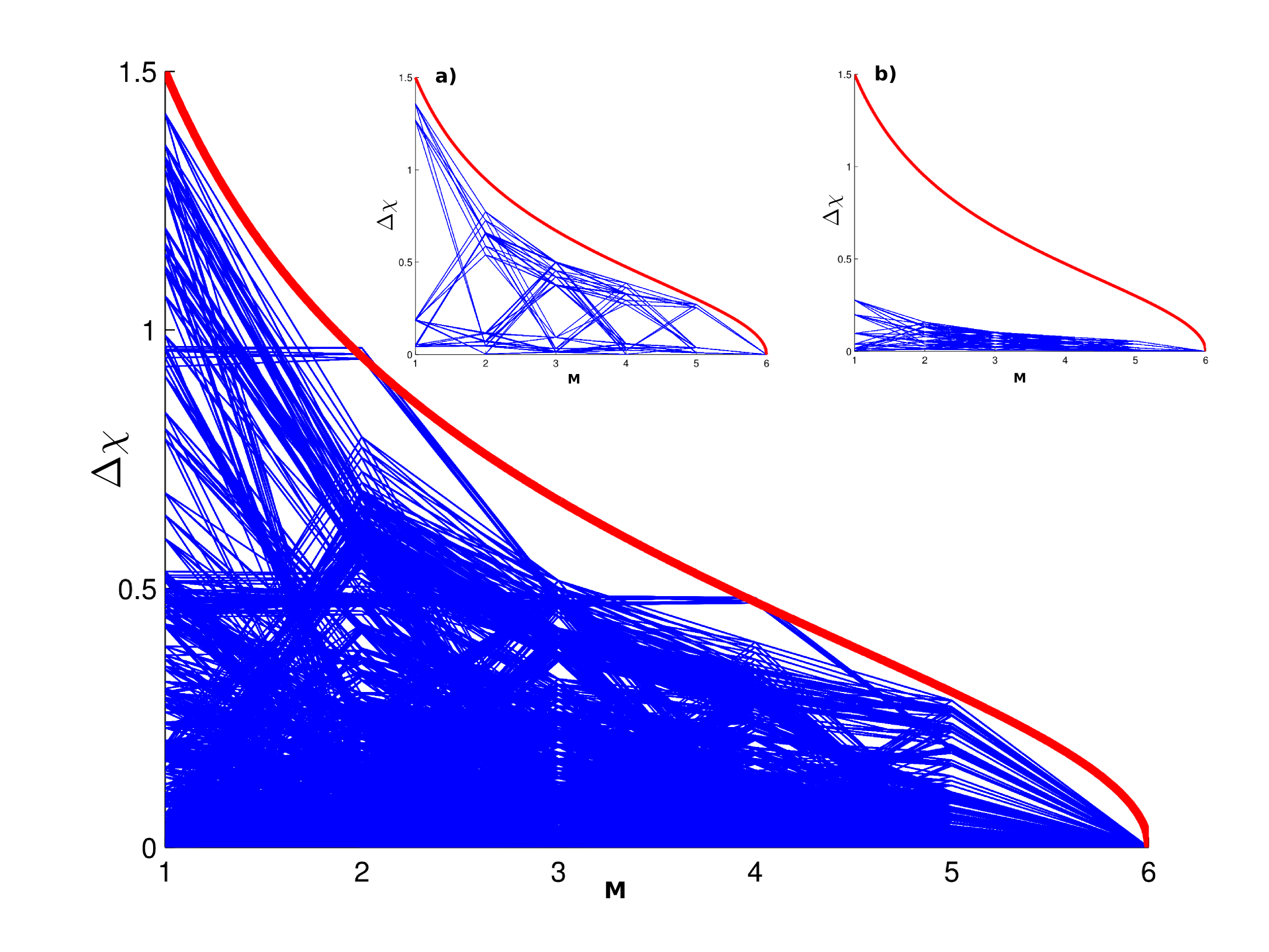, angle=0, width=0.4\textwidth}
\end{center}
\vspace{-15pt} 
\caption{Convergence for the chi matrix elements theoretical (smooth line) and experimental (segmented lines). Shown are the errors in the estimation of every $\chi_{ab}$ for both measured processes. Each curve shows the error made in the estimation for each $\chi_{ab}$ if only $M$ measurements were made. All possible choices of the first M states are plotted and they lie below the theoretical bound. The insets show the errors for two specific $\chi_{ab}$ elements. In (a) the error can stay close to the bound for some choices of the first M measurements while in (b) is always almost null.}
\label{fig:convergencia}
\end{figure}

Having the experimental data at hand we can test the behavior of the estimation error. Using the raw data we split the $2$--design in samples of different (variable) size and computed the average over each of such samples. The results are shown in Figure \ref{fig:convergencia} where we display the behavior of the estimation error for the matrix element $\chi_{ab}$ obtained by sampling the $2$--design in groups of increasing number of states (the size of the sample, $M$ grows up to $K=D(D+1)=6$, which is the cardinal of the $2$--design). These results are quite interesting and not expected: Although for a single qubit the sample space is rather small the above mentioned bound turns out to be satisfied rigorously for every instance of the experiment and not only at the statistical level. 
We see from Figure \ref{fig:convergencia} that all possible errors on all possible choices of sample partitions lie below the analytical bound. It is clear that this is not expected for all possible random distributions, but the ones realized in the experiment strictly satisfy the bound. In fact, it is not hard to imagine possible values for the results of the experiment that would violate the bound for certain samples but hold to it at the statistical level. However, such cases are not realized in the experiment which suggests that it would be possible to find a tighter bound.

We conclude by noticing that looking at the quantum state of the photon at the output of the final beam splitter in Figure 2 is quite revealing about the nature of SEQPT. In fact, it can be shown that if we prepare the initial state of the polarization qubit to be $|\psi_j\rangle$, then the unnormalized output state turns out to be described by the density matrix $\E((E_a\pm  E_b)||\psi_j\rangle\langle\psi_j|(E_a\pm E_b))$ (the $\pm$ signs are controlled by a $\phi=0,\pi$ phase with the glass in Figure 2). This shows that the ancillary (path) qubit plays a rather simple role in SEQPT: it is simply a tool to prepare a special initial state which is later sent through the channel $\E$. This observation serves as a motivation for a simple, but illuminating, reformulation of the SEQPT algorithm. In fact, SEQPT is connected with a class of quantum algorithms known as "deterministic quantum computation with one clean qubit" (DQC1) \cite{knill1998power,miquel2002interpretation}. However, as opposed to DQC1, Figure 1 shows that the measurement on the ancilla can be moved to the initial stage of the algorithm (i.e., it can be performed before the channel $\E$ acts on the system). From this point of view, SEQPT can be described as follows: The unnormalized state of the system conditioned on the detection of the $\pm 1$ eigenvalue in the measurement of $\sigma_x$ on the ancilla turns out to be $(E_a\pm E_b)|\psi_j\rangle\langle\psi_j|(E_a\pm E_b)$. Thus, after the system goes through the channel $\E$, the probability of detecting the state $\ket{\psi_j}$ is simply the fidelity of the map $\E_\pm(\rho)=\E((E_a\pm E_b)\rho(E_a\pm E_b))$. The expectation value of $\sigma_x$ is computed as the difference between the fidelities of the maps $\E_\pm$, which is nothing but the real part of $\chi_{ab}$ (analogously for the imaginary parts with $\phi=\pm\pi/2$). This new interpretation of SEQPT, suggested by the experiment indicates that the method could indeed be implemented without ancillary qubits at all. In any case, the present experiment shows the viability of SEQPT as a way to characterize a channel affecting polarization of single photons and displays its advantages over other alternative approaches. 

The experiments were performed in the quantum optics lab at CITEDEF. JPP and MAL are members of CONICET. The authors acknowledge discussions with A. Bendersky, A. Hnilo and R. Piegaia.



\begin{thebibliography}{10}%
\makeatletter
\providecommand \@ifxundefined [1]{%
 \ifx #1\undefined \expandafter \@firstoftwo
 \else \expandafter \@secondoftwo
\fi
}%
\providecommand \@ifnum [1]{%
 \ifnum #1\expandafter \@firstoftwo
 \else \expandafter \@secondoftwo
\fi
}%
\providecommand \enquote [1]{``#1''}%
\providecommand \bibnamefont  [1]{#1}%
\providecommand \bibfnamefont [1]{#1}%
\providecommand \citenamefont [1]{#1}%
\providecommand\href[0]{\@sanitize\@href}%
\providecommand\@href[1]{\endgroup\@@startlink{#1}\endgroup\@@href}%
\providecommand\@@href[1]{#1\@@endlink}%
\providecommand \@sanitize [0]{\begingroup\catcode`\&12\catcode`\#12\relax}%
\@ifxundefined \pdfoutput {\@firstoftwo}{%
 \@ifnum{\z@=\pdfoutput}{\@firstoftwo}{\@secondoftwo}%
}{%
 \providecommand\@@startlink[1]{\leavevmode}%
 \providecommand\@@endlink[0]{}%
}{%
 \providecommand\@@startlink[1]{%
  \leavevmode
  \pdfstartlink
   attr{/Border[0 0 1 ]/H/I/C[0 1 1]}%
   user{/Subtype/Link/A<</Type/Action/S/URI/URI(#1)>>}%
  \relax
 }%
 \providecommand\@@endlink[0]{\pdfendlink}%
}%
\providecommand \url  [0]{\begingroup\@sanitize \@url }%
\providecommand \@url [1]{\endgroup\@href {#1}{\urlprefix}}%
\providecommand \urlprefix [0]{URL }%
\providecommand \Eprint[0]{\href }%
\@ifxundefined \urlstyle {%
  \providecommand \doi [1]{doi:\discretionary{}{}{}#1}%
}{%
  \providecommand \doi [0]{doi:\discretionary{}{}{}\begingroup
  \urlstyle{rm}\Url }%
}%
\providecommand \doibase [0]{http://dx.doi.org/}%
\providecommand \Doi[1]{\href{\doibase#1}}%
\providecommand \bibAnnote [3]{%
  \BibitemShut{#1}%
  \begin{quotation}\noindent
    \textsc{Key:}\ #2\\\textsc{Annotation:}\ #3%
  \end{quotation}%
}%
\providecommand \bibAnnoteFile [2]{%
  \IfFileExists{#2}{\bibAnnote {#1} {#2} {\input{#2}}}{}%
}%
\providecommand \typeout [0]{\immediate \write \m@ne }%
\providecommand \selectlanguage [0]{\@gobble}%
\providecommand \bibinfo [0]{\@secondoftwo}%
\providecommand \bibfield [0]{\@secondoftwo}%
\providecommand \translation [1]{[#1]}%
\providecommand \BibitemOpen[0]{}%
\providecommand \bibitemStop [0]{}%
\providecommand \bibitemNoStop [0]{.\EOS\space}%
\providecommand \EOS [0]{\spacefactor3000\relax}%
\providecommand \BibitemShut [1]{\csname bibitem#1\endcsname}%
\bibitem{NC}%
  \BibitemOpen
  \bibfield{author}{%
  \bibinfo {author} {\bibfnamefont{M.~A.}\ \bibnamefont{Nielsen}}\ and\
  \bibinfo {author} {\bibfnamefont{I.~L.}\ \bibnamefont{Chuang}},\ }%
  \emph{\bibinfo {title} {Quantum Computation and Quantum Information}}\
  (\bibinfo {publisher} {Cambridge University Press},\ \bibinfo {year} {2000})%
  \bibAnnoteFile{NoStop}{NC}%
\bibitem{mohseni20067direct}%
  \BibitemOpen
  \bibfield{author}{%
  \bibinfo {author} {\bibfnamefont{M.}~\bibnamefont{Mohseni}}\ and\ \bibinfo
  {author} {\bibfnamefont{D.A.}~\bibnamefont{Lidar}},\ }%
  \bibfield{journal}{%
  \bibinfo {journal} {Phys. Rev. Lett.}\ }%
  \textbf{\bibinfo {volume} {97}},\ \bibinfo {pages} {170501} (\bibinfo {year}
  {2006})%
  ;
  \bibfield{journal}{%
  \bibinfo {journal} {Phys. Rev. A}\ }%
  \textbf{\bibinfo {volume} {75}},\ \bibinfo {pages} {062331} (\bibinfo {year}
  {2007})%
  \bibAnnoteFile{NoStop}{mohseni20067direct}%
\bibitem{d2001quantum}%
  \BibitemOpen
  \bibfield{author}{%
  \bibinfo {author} {\bibfnamefont{G.M.}~\bibnamefont{D'Ariano}}\ and\ \bibinfo
  {author} {\bibfnamefont{P.}~\bibnamefont{Lo~Presti}},\ }%
  \bibfield{journal}{%
  \bibinfo {journal} {Phys. Rev. Lett.}\ }%
  \textbf{\bibinfo {volume} {86}},\ \bibinfo {pages} {4195} (\bibinfo {year}
  {2001})%
  \bibAnnoteFile{NoStop}{d2001quantum}%
\bibitem{bendersky20089selective}%
  \BibitemOpen
  \bibfield{author}{%
  \bibinfo {author} {\bibfnamefont{A.}~\bibnamefont{Bendersky}}, \bibinfo
  {author} {\bibfnamefont{F.}~\bibnamefont{Pastawski}},\ and\ \bibinfo {author}
  {\bibfnamefont{J.P.}~\bibnamefont{Paz}},\ }%
  \bibfield{journal}{%
  \bibinfo {journal} {Phys. Rev. Lett.}\ }%
  \textbf{\bibinfo {volume} {100}},\ \bibinfo {pages} {190403} (\bibinfo {year}
  {2008})%
  ;
  \bibfield{journal}{%
  \bibinfo {journal} {Phys. Rev. A}\ }%
  \textbf{\bibinfo {volume} {80}},\ \bibinfo {pages} {032116} (\bibinfo {year}
  {2009})%
  \bibAnnoteFile{NoStop}{bendersky20089selective}%
\bibitem{delsarte1977spherical}%
  \BibitemOpen
  \bibfield{author}{%
  \bibinfo {author} {\bibfnamefont{P.}~\bibnamefont{Delsarte}}, \bibinfo
  {author} {\bibfnamefont{J.}~\bibnamefont{Goethals}},\ and\ \bibinfo {author}
  {\bibfnamefont{J.}~\bibnamefont{Seidel}},\ }%
  \bibfield{journal}{%
  \bibinfo {journal} {Geom. Dedicata}\ }%
  \textbf{\bibinfo {volume} {6}},\ \bibinfo {pages} {363} (\bibinfo {year}
  {1977})%
  \bibAnnoteFile{NoStop}{delsarte1977spherical}%
\bibitem{klappenecker2005mutually}%
  \BibitemOpen
  \bibfield{author}{%
  \bibinfo {author} {\bibfnamefont{A.}~\bibnamefont{Klappenecker}}\ and\
  \bibinfo {author} {\bibfnamefont{M.}~\bibnamefont{Rotteler}},\ }%
  in\ \emph{\bibinfo {booktitle} {ISIT 2005. Proc. Int. Symposium on
  Information Theory}}\ (\bibinfo {year} {2005})\ p.\ \bibinfo {pages} {1740}%
  \bibAnnoteFile{NoStop}{klappenecker2005mutually}%
\bibitem{emerson2007symmetrized}%
  \BibitemOpen
  \bibfield{author}{%
  \bibinfo {author} {\bibfnamefont{J.}~\bibnamefont{Emerson}}, \bibinfo
  {author} {\bibfnamefont{M.}~\bibnamefont{Silva}}, \bibinfo {author}
  {\bibfnamefont{et.}~\bibnamefont{al.}},\ }%
  \bibfield{journal}{%
  \bibinfo {journal} {Science}\ }%
  \textbf{\bibinfo {volume} {317}},\ \bibinfo {pages} {1893} (\bibinfo {year}
  {2007})%
  \bibAnnoteFile{NoStop}{emerson2007symmetrized}%
\bibitem{liu2008direct}%
  \BibitemOpen
  \bibfield{author}{%
  \bibinfo {author} {\bibfnamefont{W.}~\bibnamefont{Liu}}, \bibinfo {author}
  {\bibfnamefont{et.}~\bibnamefont{al.}},\ }%
  \bibfield{journal}{%
  \bibinfo {journal} {Phys. Rev. A}\ }%
  \textbf{\bibinfo {volume} {77}},\ \bibinfo {pages} {032328} (\bibinfo {year}
  {2008})%
  \bibAnnoteFile{NoStop}{liu2008direct}%
\bibitem{monz2009realization}%
  \BibitemOpen
  \bibfield{author}{%
  \bibinfo {author} {\bibfnamefont{T.}~\bibnamefont{Monz}}, \bibinfo {author}
  {\bibfnamefont{et.}~\bibnamefont{al.}},\ }%
  \bibfield{journal}{%
  \bibinfo {journal} {Phys. Rev. Lett.}\ }%
  \textbf{\bibinfo {volume} {102}},\ \bibinfo {pages} {040501} (\bibinfo {year}
  {2009})%
  \bibAnnoteFile{NoStop}{monz2009realization}%
\bibitem{altepeter2003ancilla}%
  \BibitemOpen
  \bibfield{author}{%
  \bibinfo {author} {\bibfnamefont{J.}~\bibnamefont{Altepeter}}, \bibinfo
  {author} {\bibfnamefont{et.}~\bibnamefont{al.}},\ }%
  \bibfield{journal}{%
  \bibinfo {journal} {Phys. Rev. Lett.}\ }%
  \textbf{\bibinfo {volume} {90}},\ \bibinfo {pages} {193601} (\bibinfo {year}
  {2003})%
  \bibAnnoteFile{NoStop}{altepeter2003ancilla}%
\bibitem{la_ceci}%
  \BibitemOpen
  \bibfield{author}{%
  \bibinfo {author} {\bibfnamefont{C.C.}~\bibnamefont{L\'opez}}, \bibinfo
  {author} {\bibfnamefont{B.}~\bibnamefont{L\'evi}},\bibinfo
  {author} {\bibfnamefont{D.}~\bibnamefont{G. Cory}},\ }%
  \bibfield{journal}{%
  \bibinfo {journal} {Phys. Rev. A}\ }%
  \textbf{\bibinfo {volume} {79}},\ \bibinfo {pages} {042328} (\bibinfo {year}
  {2009})%
  \bibAnnoteFile{NoStop}{la_ceci}%
\bibitem{limblad}%
  \BibitemOpen
  \bibfield{author}{%
  \bibinfo {author} {\bibfnamefont{N.}~\bibnamefont{Boulant}}, \bibinfo
  {author} {\bibfnamefont{et.}\ \bibnamefont{al}},\ }%
  \bibfield{journal}{%
  {\bibinfo {journal} {Phys. Rev. A}}\ }%
  \textbf{\bibinfo {volume} {67}},\ \bibinfo {pages} {042322} (\bibinfo {year} {2003}); %
  \bibfield{author}{%
  \bibinfo {author} {\bibfnamefont{M.}~\bibnamefont{Howard}}, \bibinfo {author}
  {\bibfnamefont{et.}~\bibnamefont{al}},\ }%
  \bibfield{journal}{%
  \bibinfo {journal} {New J. Phys.}\ }%
  \textbf{\bibinfo {volume} {8}},\ \bibinfo {pages} {33} (\bibinfo {year}
  {2006})
  \bibAnnoteFile{NoStop}{limblad}%
\bibitem{englert2001universal}%
  \BibitemOpen
  \bibfield{author}{%
  \bibinfo {author} {\bibfnamefont{B.G.}~\bibnamefont{Englert}}, \bibinfo
  {author} {\bibfnamefont{C.}~\bibnamefont{Kurtsiefer}},\ and\ \bibinfo
  {author} {\bibfnamefont{H.}~\bibnamefont{Weinfurter}},\ }%
  \bibfield{journal}{%
  \bibinfo {journal} {Phys. Rev. A}\ }%
  \textbf{\bibinfo {volume} {63}},\ \bibinfo {pages} {032303} (\bibinfo {year}
  {2001})%
  \bibAnnoteFile{NoStop}{englert2001universal}%
\bibitem{knill1998power}%
  \BibitemOpen
  \bibfield{author}{%
  \bibinfo {author} {\bibfnamefont{E.}~\bibnamefont{Knill}}\ and\ \bibinfo
  {author} {\bibfnamefont{R.}~\bibnamefont{Laflamme}},\ }%
  \bibfield{journal}{%
  \bibinfo {journal} {Phys. Rev. Lett.}\ }%
  \textbf{\bibinfo {volume} {81}},\ \bibinfo {pages} {5672} (\bibinfo {year}
  {1998})%
  \bibAnnoteFile{NoStop}{knill1998power}%
\bibitem{miquel2002interpretation}%
  \BibitemOpen
  \bibfield{author}{%
  \bibinfo {author} {\bibfnamefont{C.}~\bibnamefont{Miquel}}, \bibinfo {author}
  {\bibfnamefont{et.}~\bibnamefont{al.}},\ }%
  \bibfield{journal}{%
  \bibinfo {journal} {Nature}\ }%
  \textbf{\bibinfo {volume} {418}},\ \bibinfo {pages} {59} (\bibinfo {year}
  {2002})%
  \bibAnnoteFile{NoStop}{miquel2002interpretation}%
\end{thebibliography}

%

\end{document}